\documentclass{iau_FM}
\usepackage{graphicx}

\title[Real-time RFI Excision Techniques and their limitations] 
{Real-time RFI Excision Techniques and their limitations}

\author[Buch et al.]
{Kaushal D. Buch, $^1$
Ruta Kale $^1$, Kishor D. Naik $^1$, Ajithkumar B.$^1$, Thushara Gunaratne $^2$, N. Habana$^3$, Gregory Hellbourg $^4$, Jane Kaczmarek $^5$, L. Petrov $^3$, Cedric Viou $^6$, Benjamin Winkel $^7$ 
} 

\affiliation{$^1$Giant Metrewave Radio Telescope, NCRA-TIFR, \\ Pune 410504,
Maharashtra, India \\ email: {\tt kdbuch@gmrt.ncra.tifr.res.in, ruta@ncra.tifr.res.in, kishornaik33@gmail.com, ajit@ncra.tifr.res.in} \\[\affilskip]
$^2$Herzberg Astronomy and Astrophysics Research Center, \\ National Research Council Canada \\ 
Canada \\email: {\tt Thushara.Gunaratne@nrc-cnrc.gc.ca} \\
$^3$ NASA Goddard Space Flight Centre, \\ Maryland, USA \\ mail: {\tt n.habana@nasa.gov, leonid.petrov-1@nasa.gov} \\
$^4$ California Institute of Technology, \\
Pasadena, CA, USA \\email: {\tt ghellbourg@astro.caltech.edu} \\
$^5$ CSIRO Space \& Astronomy, Australia \\
email: {\tt jane.kaczmarek@csiro.au} \\
$^6$ Nancay Observatory, France \\
email: {\tt Cedric.Viou@obs-nancay.fr} \\
$^7$ Max Planck Institute for Radio Astronomy (MPIfR), \\
Bonn, Germany \\ email: {\tt bwinkel@mpifr-bonn.mpg.de}
}
\pubyear{2024}
\jname{Astronomy in Focus, Focus Meeting 5} 
\editors{}
\begin{document}

\maketitle

\firstsection 
\section{An Overview of Real-time RFI Excision Techniques}
Contemporary real-time RFI mitigation is carried out at different stages primarily using regulatory and technical approaches. Regulatory approaches include spectrum management, radio quiet zones, and ensuring protection from self-generated RFI. The technical approaches include mitigation RFI in the analog and RF frontend systems, digital signal processing systems, and offline systems. 

As is known, the signal received by a radio telescope is a combination of the contributions from the astronomical signal and a combination of system and sky background noise. RFI has an additive effect on the signal received by the radio telescope. The key distinguishing properties of RFI are that it is generally stronger than the signal and non-Gaussian. 

The capability of signal processing receiver systems has grown manifold with the advent of high-speed signal processing platforms like Field Programmable Gate Arrays (FPGA) and Graphics Processing Unit (GPU). This has enabled the development of different signal-processing techniques for real-time RFI mitigation algorithms. Primarily these are divided into the following three categories.

\begin{itemize}
\item Excision: This technique removes strong RFI through statistical methods and is relatively easier to implement in real-time. However, it leads to a larger data loss. The technique is widely used and can be applied to time and frequency domains.
\item Cancellation: A reference antenna samples the RFI and adaptively subtracts the RFI from the combination of signal+RFI. This is complicated to implement but has a lower data loss than Excision.
\item Nulling: This technique works well for telescopes with phased array feeds and cases where the RFI source is within the field of view. This technique involved beamforming with multiple feed elements and hence is difficult to implement.
 \end{itemize}
 
The techniques that can be used depend on the type and source of RFI which also decide the mitigation location in the receiver system.
\subsection{Upgraded GMRT (uGMRT) Case Study}
Giant Metrewave Radio Telescope located 80 km north of Pune in India is an array of thirty, 45m, fully steerable parabolic dishes. The array is spread over an area of 25km.  It has been recently upgraded to provide near-seamless coverage from 120 MHz to 1450 MHz, with a maximum instantaneous bandwidth of 400 MHz. uGMRT is affected by broadband and narrowband RFI from various sources like powerlines, communication transmitters, broadcast services, and navigational satellites, to name a few. 

Here, we describe the real-time RFI mitigation system developed and commissioned at the GMRT to mitigate broadband RFI from powerlines. 
Since powerline RFI is broadband, frequency selective filtering in the receiver system cannot be used for mitigation. Hence, we use statistical threshold-based outlier detection and filtering in real-time on digitized time series from each antenna, in the pre-correlation domain. 

The algorithm uses three steps - Estimation, Detection, and Filtering. Robust estimation of the dispersion of the signal is carried out using Median Absolute Deviation (MAD) and its variant. This is followed by threshold estimation and comparison with the threshold values. Any sample outside the threshold is detected as RFI, replaced by a user-defined value, and a flag is generated corresponding to that sample. 
Several architectural and hardware optimized carried out are described in \cite{Buch_etal19}. 

We came up with a concept of simultaneous testing wherein half the antennas pass through the filtering system and their copies without the filtering system. 
In the simultaneous testing mode, the imaging experiments showed about 3 dB improvement in the noise RMS, particularly on shorter baselines ($<$0.5 km) with an average filtering of 2-3 \% data. Similarly, about 3 times improvement was seen in the SNR for time-domain experimental observation.

After rigorous testing, we released the system (full release, April 2022) for use during the regular GMRT observations. 
In the last observing cycle, about 520 hours of observing was carried out, which is about 30\% of the total observing time. 

Currently, we are involved in the analysis of the filtering scheme on visibility data, developing an algorithm for narrowband RFI mitigation, porting it on different platforms like CPU/GPUs, extending the algorithm \cite{Buch_etal23} to the Expanded GMRT \& SKA LFAA project and incorporating learning-based approaches.

\section{RFI Mitigation Strategies at Murriyang, the CSIRO Parkes Radio Telescope}
CSIRO Parkes telescope is a single-dish telescope with a 64m diameter parabolic dish antenna. It is the largest single-dish telescope in the Southern Hemisphere. 
It has been continuously upgraded and will provide frequency coverage from 0.7 to 26 GHz. The future receiver fleet consists of three main receiver systems - UWL, CryoPAF, and UWM-H. These systems are outstanding, always available, and have a continuous frequency coverage. However, due to its wide bandwidth, and in the case of PAF, the wide field of view, they are susceptible to RFI. 

In terms of the signal processing capabilities, the Parkes has one backend end i.e. Medusa to take care of all the receiver systems mentioned above. Medusa is a GPU-based, highly flexible backend, with user-defined sampling rate, frequency resolution, phase binning, and support for multi-modal observing. 

The current RFI mitigation strategy includes excision using the spectral kurtosis approach which is implemented in quasi real-time on GPU. The algorithm detects non-gaussianity in the data and has been shown to successfully mitigate the RFI from WiFi. Also, this has been demonstrated on a pulsar observation.

The cryogenically cooled PAF on the Parkes telescopes consists of 72, dual-polarisation, steerable beams. 
This will lead to a 10-30x increase in survey speed and ~2 deg2 field-of-view. Spatial nulling (or beam nulling) using this system has been demonstrated on an observation for mitigation RFI in the L2 band at 1225 MHz (with a bandwidth of 1 MHz). The mitigation on and off plots show a large improvement due to the nulling process \cite{Hellbourg_et_al16}. Similarly, improvement is seen using beam nulling for the RFI from GPS satellites in the L5 band (1175 MHz). The technique works for a moving and stationary source of RFI.

\section{On the Alleviation of the Impact of Spaceborne RFI on VLBI Results}

VLBI observes extragalactic sources. Their emission is much fainter than the emission from artificial satellites. Interference from satellite ranges from mild when a portion of the bandwidth has to be flagged out, to strong when a receiver is driven to the saturation mode and all the data from a given source is lost, and catastrophic when satellite emission causes permanent damage to the frontend. An increase in the number of communication satellites and their transmitted power aggravates the problem. The first line of defence is to
make a schedule in such a way that we do not point the antenna in directions in the vicinity of those satellites that may cause interference. This is achieved by generating
station- and time-dependent dynamic masks that define the zone of avoidance and using them during schedule preparation.

We built a database with satellite catalogues containing information about the various parameters. It includes two-line element ephemeride of each satellite are used for projecting the trajectories to compute the look angles. 
The threat level is determined based on FCC filing and look angles. 

To demonstrate this, we set up a 0.6m filed antenna at the Goddard Geophysical and Astronomical Observatory (GGAO) at Greenbelt, Maryland, USA. This antenna is colocated with the 12m dish and has a wide beam and large bandwidth (2-18 GHz).  The results from dynamic masking are encouraging and it was tried for a few satellites. 

The future scope of this work includes synchronizing this antenna with the 12m dish, incorporating the dynamic mask in the VLBI observing schedule, and developing a website containing information about the dynamic mask.

\end{document}